\font\twelverm = cmr10 scaled\magstep1 \font\tenrm = cmr10
       
\font\twelvei = cmmi10 scaled\magstep1
       \font\teni = cmmi10 
\font\twelveit = cmti10 scaled\magstep1 
       
\font\twelvesy = cmsy10 scaled\magstep1
       \font\tensy = cmsy10 
\font\twelvebf = cmbx10 scaled\magstep1 \font\tenbf = cmbx10
       
\font\twelvesl = cmsl10 scaled\magstep1
\font\twelveit = cmti10 scaled\magstep1
\font\twelvett = cmtt10 scaled\magstep1
%
\textfont0 = \twelverm \scriptfont0 = \twelverm
       \scriptscriptfont0 = \tenrm
       \def\rm{\fam0 \twelverm}
\textfont1 = \twelvei \scriptfont1 = \twelvei
       \scriptscriptfont1 = \teni
       
\textfont2 = \twelvesy \scriptfont2 = \twelvesy
       \scriptscriptfont2 = \tensy
       
\newfam\itfam \def\it{\fam\itfam \twelveit} \textfont\itfam=\twelveit
\newfam\slfam  \textfont\slfam=\twelvesl
\newfam\bffam \def\bf{\fam\bffam \twelvebf} \textfont\bffam=\twelvebf
       \scriptfont\bffam=\twelvebf \scriptscriptfont\bffam=\tenbf
\newfam\ttfam  \textfont\ttfam=\twelvett
\rm
\hsize=6in
\hoffset=.45in
\vsize=9in
\baselineskip=24pt
%
\raggedright  \pretolerance = 800  \tolerance = 1100
\raggedbottom
%
\dimen1=\baselineskip \multiply\dimen1 by 3 \divide\dimen1 by 4
\dimen2=\dimen1 \divide\dimen2 by 2
%
\nopagenumbers
\headline={\ifnum\pageno=1 \hss\thinspace\hss
     \else\hss\folio\hss \fi}
%
\def\heading#1{\vfill\eject \vbox to \dimen1 {\vfill}
     \centerline{\bf #1}
     \vskip \dimen1}
%
\newcount\sectcount
\newcount\subcount
\newcount\subsubcount
\global\sectcount=0
\global\subcount=0
\global\subsubcount=0
\def\section#1{\vfill\eject \vbox to \dimen1 {\vfill}
    \global\advance\sectcount by 1
    \centerline{\bf \the\sectcount.\ \ {#1}}
    \global\subcount=0  
    \global\subsubcount=0  
    \vskip \dimen1}
%
\def\subsection#1{\global\advance\subcount by 1
    \vskip \parskip  \vskip \dimen2
    \centerline{{\it \the\sectcount.\the\subcount.\ \ #1}}
    \global\subsubcount=0  
    \vskip \dimen2}
%
\def\subsubsection#1{\global\advance\subsubcount by 1
    \vskip \parskip  \vskip \dimen2
    \centerline{{\it \the\sectcount.\the\subcount.\the\subsubcount.\ \ #1}}
    \vskip \dimen2}
%
%
\def\refindent{\advance\leftskip by 24pt \parindent=-24pt}
%
\def\journal#1#2#3#4#5{{\refindent
                      {#1}        
                      {#2},       
                      {#3},       
                      {#4},       
                      {#5}        
                      \par }}
%
\def\infuture#1#2#3#4{{\refindent
                  {#1}         
                  {#2},        
                  {#3},        
                  {#4}         
                  \par }}
%
\def\inbook#1#2#3#4#5#6#7{{\refindent
                         {#1}         
                         {#2},        
                      in {\it #3\/},  
                     ed. {#4}         
                        ({#5}:        
                         {#6}),       
                       p.{#7}         
                         \par }}
%

%
\def\book#1#2#3#4#5{{\refindent
                   {#1}         
                   {#2},        
                   {\it #3\/}   
                  ({#4}:        
                   {#5})        
                   \par }}
%
\def\privcom#1#2#3{{\refindent
                  {#1}        
                  {#2},       
                  {#3}        
                  \par }}
%

%

%

%

\def\sun{_\odot}
\def\samename{\vrule height0.4pt depth0.0pt width1.0in \thinspace.}
%
%
%

\newcount\notenumber
\notenumber=0
\def\note{\global\advance\notenumber by 1
	  \footnote{$^{\the\notenumber}$}}
\def\apjnote#1{\global\advance \notenumber by 1
	       $^{\the\notenumber}$ \kern -.6em
               \vadjust{\midinsert
                        \hbox to \hsize{\hrulefill}
                        $^{\the\notenumber}${#1} \hfill \break
                        \hbox to \hsize{\hrulefill}
                        \endinsert
                       }
              }
\hyphenpenalty=200
\
\vskip 3cm
\centerline{ON THE ORIGIN OF THE GALACTIC MAGNETIC FIELD}
\vskip 2cm
\settabs 3 \columns
\+&Ethan T. Vishniac\cr
\+&Dept. of Astronomy\cr
\+&University of Texas\cr
\+&Austin, TX 78712\cr
\vskip 10cm
\line{Received:\hrulefill}
\bf
\heading{Abstract}
The galactic magnetic field is commonly supposed to be due to a
dynamo acting on some large scale seed field.  A major difficulty
with this idea is that estimates of reasonable seed field strengths
tend to be quite low, on the order of $\sim10^{-20}$ gauss.
Here we examine the contribution due to the flux entrained
in winds from protostars formed in the first dynamo e-folding time
of a galaxy's existence.  Using a minimal estimate of a protostellar
magnetic field we find that if each protostar ejects
a single current ring, sufficient to maintain flux freezing in
the wind, than the large scale average dipole field from all such
current rings will be at least 5 orders of magnitude larger than
previous seed field estimates.  Allowing for a reasonable amount
of magnetic activity in protostars during an extended period of
mass loss increases this to a dipole seed field of $\sim10^{-12}$ gauss.
For the purposes of producing a seed field it is irrelevant whether or
not this initial injection of flux takes place in a newly formed
galactic disk, or in star forming proto-galactic clouds.  The compression
of this dipole field into a thin disk will lead to a large scale
$B_r\sim 10^{-10.5}$ gauss.  Initially, field strengths on smaller scales
will be larger, but nowhere near current levels.
\rm
\vfill
\eject
\section{INTRODUCTION}

The origin of galactic magnetic fields is a long standing problem in
theoretical astrophysics.  Virtually all spiral galaxies have magnetic
fields of a few microgauss which are spatially coherent over several
kiloparsecs.  This implies that the average magnetic energy density
is comparable to the average energy densities of cosmic rays and the
interstellar medium, which in turn suggests that galactic magnetic fields
reached
a state of saturation some (unknown) time in the past.  Since galaxies
are so large, and their constituent gases are so highly conducting, the
dissipative time scale for these fields is absurdly long.  Either galactic
magnetic fields are simply the visible manifestation of some pre-existing
cosmological field (cf. Piddington 1972; Kulsrud 1990) or they are spectacular
examples of
fast dynamos, in which turbulent transport takes the place of
ohmic dissipation (Parker 1971; Vainshtein and Ruzmaikin 1971, 1972).
The notion that there exist large scale
magnetic fields as a consequence of physical processes that
occur in the very early universe has been explored by a number of
authors (Vilenkin \& Leahy 1982; Hogan 1983; Turner \& Widrow 1988;
Quashnock, Loeb \& Spergel 1988; Vachaspati 1991; Ratra 1992; Dolgov \& Silk
1992
Dolgov \& Rhie 1992).  The bulk of this work is neither clearly wrong,
nor clearly right, inasmuch as it requires the invention of novel
physical mechanisms for which no evidence can be found (aside from
the existence of galactic magnetic fields themselves).  A few, such as
Quashnock
et al. (1988), suggest mechanisms which may generate small
scale fields, but which cannot, by themselves, give rise to large scale
coherent
fields.  One of the earliest (Harrison 1970) pointed out that such fields
would arise naturally from large scale rotational motions during the radiation
epoch,
but the existence of such motions are not easily reconciled with standard
cosmological models and constraints on their amplitude limit the resulting
field
to very low levels.

One is therefore tempted to search for the source of the field among
physical processes that happen during, or after, the epoch of galaxy
formation, when the relevant physics can be assumed to be
testable, if not completely understood.  In this picture one assumes
that the galactic field grows from some small seed due to the action
of an $\alpha-\Omega$ dynamo, i.e. a dynamo in which some small mean
helicity generates a radial magnetic field, $B_r$, from an azimuthal
one, $B_\theta$.  Differential rotation in the galactic disk then closes
the loop by generating $B_\theta$ from $B_r$.  The growth rate is roughly
$B_r/B_\theta\sim(\alpha_{\theta\theta}\Omega/H)^{1/2}$,
where $\alpha_{\theta\theta}$ is the azimuthal component of the helicity
tensor, $H$ is the disk scale height, and $\Omega$ is the disk rotation
frequency (about $10^{-15}$ in the solar neighborhood).  There is no
general agreement on the source, or value, of $\alpha_{\theta\theta}$
but given the largely azimuthal nature of galactic magnetic fields
it seems reasonable to assume that the dynamo growth rate ought to
be $\sim 10^{-16}$ to within factors of order unity.  Given a current
field strength of $3\times 10^{-6}$ gauss one finds that a seed field
of $\sim 3\times 10^{-19}$ gauss should be sufficient to produce the
observed fields after 10 billion years (but see below!).  Of course,
the uncertainties in growth times and
the ages of galactic disks enter into the exponent of this calculation,
so the result cannot be taken too literally.

Can such a seed field be produced?  One can, of course, return to the
early universe as a source, with the advantage that one requires a much
weaker cosmological effect, and therefore a much smaller modern intergalactic
field, but the problem remains that such models are only loosely
connected to testable physics.  There are two proposals which involve
the direct creation of a large scale galactic field using only standard
physics, the microwave background compton drag on electrons in the
galactic disk (Mishustin and Ruzmaikin 1971), and the thermoelectric effect
(Biermann 1950; Roxburgh 1966; Lazarian 1992).  The former is due to the
deceleration of the electrons
in a rotating galactic disk due to their interaction with the microwave
background.  This creates a large scale dipole field at a rate of
$$\partial_t\vec B={m_pc\over e}{\vec\Omega\over \tau_{e\gamma}} \eqno(1)$$
$m_p$ is the proton mass, $c$ is the speed of light, $e$
is the charge of an electron, $\Omega$ is the rotation rate of the galaxy,
and $\tau_{e\gamma}$ is the electron-photon coupling time.  This field
will begin to grow as soon as the protogalactic disk forms.
Since $\tau_{e\gamma}\propto (1+z)^{-4}$ we note that most of the field
creation takes
place shortly after galaxy
formation.  Plugging in a galactic rotation rate of $10^{-15}$, a
galaxy formation epoch of $z=5$ in a flat cosmological model,
and a current background temperature
of $2.75$ K  we find a seed field
of $1.6\times10^{-20}h^{-1}$ gauss, where $h$ is Hubble's constant in
units of 100 km/s/Mpc.  This result scales with the redshift of galaxy
formation as $(1+z)^{5/2}$ and the value we have used is, if anything, a little
high, so this mechanism would appear to be marginal at best.

The thermoelectric effect comes from a misalignment of temperature
and density gradients in a gas, a circumstance that would naturally
arise in a rotationally supported system.  The field grows linearly
in time, at a rate of
$$\partial_t\vec B\approx{m_p c\over e}{\vec\nabla
P{\bf\times}\vec\nabla\rho\over \rho^2}\eqno(2)$$
For a rotationally supported disk this is approximately
$$\partial_t\vec B\sim {m_p c\over e} {H\over r} \Omega^2\eqno(3)$$
where $H$ is the disk thickness and $r$ is the disk radius.
We note that smaller systems have a tremendous advantage in generating
magnetic fields in this way since $\Omega$ will scale as $H^{-1}$ for
similar temperatures, a point previously stressed by Pudritz \& Silk (1989).
If we follow Lazarian (1992) and take
a gas temperature of $10^6$ K in a disk with a radius of 10 Kpc
we find that this mechanism will produce a galactic seed field of
approximately $3\times 10^{-19}$ gauss in $10^{16}$ seconds.  This seems more
promising, although still somewhat marginal, assuming a dynamo growth time
comparable to the current age of our galaxy. (Lazarian quotes a value of
$3\times10^{-17}$
gauss, but the discrepancy is due to an integration time of $10^9$ years and a
vertical
scale height for the hot gas of $100$ pc.  The former parameter corresponds to
an
integration time of at least 3 dynamo e-folding times, which seems unnecessary.
 The
scale height is unrealistically small.  We take $H\sim c_s/\Omega$.)
However, such disagreements are unimportant.  The real problem
with this proposal has to do with the time required
for the galactic dynamo to reach saturation.

Recent work on inferred magnetic field strengths for high redshift systems
(Kronberg, \& Perry 1982; Welter, Perry, \& Kronberg 1984; Wolfe, Lanzetta, \&
Oren 1992)
has dramatically shortened the time available for galactic
dynamos to work.  Kronberg and collaborators initially showed that
observations of Faraday rotation measures of QSOs with,
and without, metal line absorption systems, indicates that dynamically
significant magnetic fields were already present at redshifts of a few.
Since then Wolfe et al. showed that this effect was strongest for QSOs with
Lyman limit systems,
and that these systems, usually identified with early galactic disks,
contain kiloparsec scale coherent magnetic fields with strengths of a few
microgauss
at redshifts of about 2.  The strength and scale of these fields
are indistinguishable from the fields of galaxies today, indicating that
galactic
dynamos reach saturation in a short time, not much more than a billion years.
Assuming that galactic dynamos grow at less than the galactic rotation rate
(typically
about $10^{-15}\ s^{-1}$) this implies a total amplification less than about
$10^{19.5}$.
More realistic galactic dynamos, with growth rates of $\sim 10^{-16}$, cannot
possibly
amplify a seed field of $\sim 10^{-20}$ gauss to saturation in such a
short time.  This has led to a renewed interest in the early universe as a site
for seed field generation.

Is it really necessary to resort to such speculative processes?
Here we will argue that the process of star formation itself will
inevitably lead to a significant galactic seed field, one that is
larger than that contributed by any but the most ad hoc cosmological
processes.  We start by considering current observations of very young
stellar objects.  It is well known that such objects appear to drive
very strong molecular outflows and that such outflows accompany the formation
of comparatively low mass stars (for a general review see Fukui et al. 1992
and references contained therein).  These outflows are broad
(length-to-width ratios of 3 to 1 appear to be typical), large (with typical
dimensions of $\sim 10^{18}$ cm) and dense (with typical densities of
about $10^3 cm^{-3}$).  In addition to these broad outflows, protostars have
associated optical jets, on scales of $\sim 10^{17}$ cm.  These jets
were discovered only recently (Dopita, Schwartz, \& Evans 1982; Mundt \& Fried
1983) but are probably the source of the momentum contained in the broader
molecular
outflows (Stahler 1992).

Such outflows are direct evidence that protostars
can have a dramatic influence on their environment.  The outflows observed near
protostars are sufficiently energetic that their combined effects would suffice
to drive gas out of star forming regions altogether, even in the absence
of extremely massive and luminous stars.  Here we will argue
that these flows should also contain a dynamically insignificant poloidal
field which, when summed over an entire galaxy undergoing its first burst
of star formation, will lead to a large galactic seed field.
This process is analogous to the way that the solar wind transports the
poloidal field of the Sun outward, but the magnetic fields and radii of
protostars are much larger, so the affect will be much greater.
We will start by considering the minimal plausible value for the
magnetic field of a protostar driving these intense outflows.  Once we have
obtained this estimate we can consider whether such fields are likely to
be entrained in the mass outflow and what the total effect of many such star
forming regions is apt to be.

\section{ESTIMATING THE MINIMUM GALACTIC SEED FIELD}

Let's start by considering a protostar embedded in a disk.  The bulk
of the disk is apt to be too dense and cool to contain a significant
ionized component (Hayashi 1981; Umebayashi 1983; Umebayashi \& Nakano 1988)
although there are some indications that grains
in the inner solar system were exposed to magnetic fields on the order of
a gauss or so (Levy \& Sonett 1978).  Nevertheless, it is clear that  close to
the protostar
the ionization fraction will be high and the gas will be a good conductor.
Since the thermoelectric effect will be much more effective on small
scales than on large ones, we expect the protostar to have a significant
seed field, even if it forms in the absence of a strong galactic field.
Moreover, we know that stars have strong internal dynamos, although we do
not have a detailed understanding of their physical basis.  One final
point is that viable theories of jet formation all rely on some sort of
magnetic driving mechanism.  Taken together, these facts imply that
protostars generate internal magnetic fields which are eventually
strong enough to drive the dramatic outflows we see today.  Since
we are interested in deriving the minimal poloidal magnetic flux that should
be imprinted on the outflow we need to know the minimum stellar field
that could be expected to strongly bias the internal field of
the surrounding disk, and the minimum stellar field that
could be expected to drive outflows of the observed strength.

We will begin with the first question.  This is complicated by the fact that
the disk itself may contain a dynamo mechanism which might dominate
its internal poloidal field.  How can we be sure that the stellar
field dominates?  One simple test follows from considering the stability
of a magnetic field in an accretion disk.  A vertical field threading
a conducting disk will be unstable (Velikhov 1959; Chandrasekhar 1961; Balbus
and Hawley
1991; Hawley and Balbus 1991).  The consequent turbulence will
induce local angular momentum transport equivalent
to a dimensionless viscosity of $\alpha\sim V_A/c_s$, where $V_A$
is the Alfv\'en speed associated with the imposed field and $c_s$ is
the local sound speed (Vishniac \& Diamond 1992; Liang, Diamond \& Vishniac
1993).  Clearly this
field will dominate the internal structure of the disk, and resist
distortion by some internal disk dynamo, provided that
$V_A\sim c_s> \alpha_d$ where $\alpha_d$ is the internally generated
value of $\alpha$.  This implies that there is a critical field strength
for the protostar given by
$$B_{z,c}\sim \alpha_d (4\pi P)^{1/2},\eqno(4)$$
where $P$ is the pressure in the disk, evaluated near the inner edge of the
disk.
Now in an $\alpha$ model disk (Shakura \& Sunyaev 1973) the mass flow through
the disk is approximately
$$\dot M\sim 3\pi \alpha_d {PH\over \Omega}\eqno(5)$$
Therefore we can rewrite Eq. (4), dropping factors of order unity, as
$$\eqalignno{B_{z,c}&\sim \left({\alpha \dot M\Omega\over
H}\right)^{1/2}&(6)\cr
&\sim \left({\alpha r\over H}\right)^{1/2}\left({\dot M\Omega\over
r}\right)^{1/2}}$$
where all these quantities are evaluated at the inner edge of the disk.
In general the leading factor of $\alpha r/H$ will be of order unity or less.
Wood \& Mineshige (1989)
have shown on phenomenological grounds that in cataclysmic variable disks
$\alpha$ scales as
$(H/r)$ to some power close to one.  Angular momentum transport in low mass
protostellar disks is
unlikely to be more efficient than in fully ionized systems.  It may be less
efficient.

What protostellar magnetic field will be sufficient to drive significant mass
loss from the disk?  Observed jets in these systems have relatively high mass
fluxes and
involve material flowing outward at hundreds of km/sec.  Evidently the outward
mass flux
is comparable to the inward mass flux within the disk.  Somehow a large
fraction
of the mass contained in the inner edge of the disk is diverted into a
wind.  The stellar magnetic field required to drive such an outflow is
uncertain,
but a model-independent minimal condition can be derived.
The torque per unit area exerted by a dipole field that penetrates the
disk is $B_\theta B_z r/4\pi$.  Assuming that the dipole field acquires
a distortion of order unity (which should be about right if the star is
rotating rapidly and reconnection occurs in the disk atmosphere) then
this becomes critical when
$${B^2 r\over 4\pi \Sigma}=\alpha H^2\Omega^2\eqno(7)$$
or
$$B_{z,c}\sim \left({\dot M\Omega\over r}\right)^{1/2}\eqno(8)$$
For field strengths this large the torque exerted on a typical fluid
element in the disk will be more important than the angular momentum
transport due to processes internal to the disk.  Large amounts of
material will spiral outward, eventually turning into an intense
outflow (as observed!).  We don't actually know how protostellar magnetic
fields grow or saturate, but some kind of dynamo process should set in at the
bottom of the Hayashi track, i.e.  when a star is no longer fully convective.
For `typical' protostar values, i.e. $\dot M\sim 3\times 10^{19}$ gm/sec,
$r\sim 10^{12}$ cm, $r\Omega\sim 10^7$ cm/s, this gives a critical
field strength of about 20 gauss.  There is nothing unreasonable about
a magnetic field of this strength.  In fact, observational studies of T Tauri
stars (Bouvier \& Bertout 1989) suggest that as much as $10\%$ of the surface
is covered with photospheric spots with a typical local flux of 1300 gauss.
Real fields might exceed the critical limit we have derived here by more than
an order of magnitude.  Work by Saar \& Linsky (1985) and Basri \& Marcy (1988)
confirms that such field strengths are not unusual for late-type active stars.
For comparison we note that the Sun has an average dipole field of a few gauss.
In what follows we will take a conservative line and invoke only the minimal
field estimate
derived above.

Since $\alpha\sim H/r$ or less, a magnetic field strong enough to drive a
strong
outflow will necessarily be strong enough to penetrate into the inner regions
of the surrounding accretion disk.  This implies that the outflow will contain
the
current induced in the disk by the protostellar magnetic field. Moreover,
since the outflow is composed of a hot, and well conducting, gas we expect
that this current will tend to maintain the flux threading the outflow as it
moves away from the star.
The net affect will be to transport a flux, typical of the protostar
at the epoch of outflow, out to large distances.  Eventually this flux
will be spread out over distances comparable to, or greater than, typical
stellar separations.  The strength of the field at this point depends on
the number of separate current rings carried by the outflow.
If the protostar undergoes $N_m$ field reversals while ejecting matter,
then outflow will contain $N_m$ current rings, which will add incoherently
to produce the total flux.  Assuming the protostar reverses on a time
scale of years, and the outflow persists for a few million years,
this suggests an additional factor of $10^3$.  Observationally not much
is known about the frequency of magnetic field reversals in such young stars,
but their rotation periods are known to be short (on the order of a day)
and one might expect on theoretical grounds (Parker 1955, 1979; for a recent
treatment
see Stix 1989) that this would indicate
a proportionately shorter magnetic reversal time scale.

The total large scale field for the whole galaxy
that is generated from the incoherent addition of the fields generated
by $N$ newly formed stars will be
$$B_{seed}\sim \left({N N_m\dot M(\Omega r)r^2\over
r_g^4}\right)^{1/2}\eqno(9)$$
where $r_g$ is the radius of the galaxy.  In other words, the total large
scale flux will be given by the incoherent addition of the contributing flux
elements.  In more physical terms this is due to the aggregation of the ejected
current rings as the mass outflows from various protostars encounter one
another.
The star formation rate when the
galaxy was young is not known precisely, but was probably at least
30 stars per year.  Taking a dynamo growth rate of a few times $10^{-16}
sec^{-1}$
this implies $N\sim 3\times 10^9$.  Then for $r_g\sim 3\times 10^{22}$ cm we
get
$$B_{seed}\sim 10^{-12} gauss\eqno(10)$$
which is 7-8 orders of magnitude larger than earlier suggestions based
on known physics.  The flattening of
the poloidal field due to its confinement in the disk will produce a further
amplification of $B_r$ of about $30$ so our final result is a minimal
large scale field strength of $\sim 3\times 10^{-11}$ gauss.

We have assumed that the flux estimate should be made for a field strong
enough to disrupt the disk at its inner edge.  However, the field
itself does not depend on input from the accretion disk and so will
persist even after the disk is eroded.  Moreover there is some evidence
that at late times there may be some expulsion of mass from the protoplanetary
disk even at rather large radii.  For example, in our own solar system
the gas of the protostellar disk appears to have been expelled out to well
beyond the orbit of Mars.  This may imply that young stars can support
stronger magnetic fields, and therefore eject substantially more
flux than the estimates given here.  On the other hand, it may simply
indicate that at late times the internal processes supporting the
transfer of angular momentum within accretion disks become weak
and even relatively modest fields can lead to a deep erosion of the
protoplanetary disk.

We note in passing that this hypothesis depends on the advected flux acting
passively,
i.e. that separate current rings do not exert significant forces on one
another.
One way of asking whether or not this is reasonable is to evaluate the typical
field
strength on scales that would typically separate protostars.  If the associated
magnetic field
pressure is large compared to typical ISM energy densities then we have
exaggerated the
ability of protostellar outflows to seed the galactic magnetic field.  For the
numbers
taken above and assuming the protostars are distributed evenly within the
galactic disk,
we find a typical separation of $\sim 5$ pc and a typical field strength on
that scale
of $B_z\sim 10^{-10}$ gauss.  This estimate ignores the amplification of $B_r$
that comes from
the flattening of the disk.  If that affects all scales equally then this
estimate needs to
be raised to $\sim 10^{-8.5}$ gauss.  Of course, protostars are not distributed
evenly at present,
and probably weren't at earlier epochs, but the modest size of this estimate
suggests that
local field required in this picture is dynamically insignificant.

\section{CONCLUSIONS}

It is disappointing to note that our results here are extremely approximate.
A more precise estimate would require a detailed and plausible model for the
generation
of the outflow, a complete knowledge of the evolution of the disk and protostar
as a function of the total mass, the initial mass function in the galaxy and a
detailed understanding of the
protostellar dynamo and the stability of the protostellar magnetic field.
Nevertheless, the
crude estimate we have arrived at here is sufficient to draw the following
conclusions.
First, the ejection of current rings from protostellar systems produces a
galactic
seed field with $B_r\sim 10^{-10.5}$ gauss.  This is much larger than any other
physically realistic seeding mechanism.  Second, the generation of the seed
field does not depend on the detailed history of the galaxy.  This seeding
process works equally
well if the galaxy begins as a loosely associated cloud of star forming
regions.
Third, we still require a fast dynamo with a growth rate slightly more than
$10^{-16} s^{-1}$
to grow microgauss fields in a little more than a billion years.  The large
uncertainties in the seeding process can only change this estimate by some
logarithmic
factor.  We note that the existence of galactic dynamo is still
somewhat controversial. Kulsrud (1990) has suggested that any such dynamo
should
overproduce small scale fields.  Since his calculation assumed that such fields
were essentially passive this objection may not be realistic.  On the other
hand,
Vainshtein \& Cattaneo (1992) have posed an objection, in principle, to {\it
any} fast dynamo process.
Since real astrophysical systems would be difficult to understand if fast,
large
scale, dynamos were actually impossible, it seems likely that nature has some
way
of evading this objection (for one such suggestion see Vainshtein, Parker, \&
Rosner 1993).  Here we
simply note that this work assumes such a dynamo is possible.  If not, then our
seeding mechanism is insufficient to explain galactic magnetic fields.

It is interesting to compare this process to one based on flux ejection
from collapsed objects.  There are at least two basic difficulties
encountered in trying to apply a similar idea to black holes.
First of all, the energy required for escape from a black hole is
such that subsequent escape from the galaxy
can be avoided only if the jet plows into a large column density of
interstellar gas.  This implies that large black holes may produce
jets, as seen in AGN, but will have some difficulty in mixing ejected
flux into the galactic disk and halo.  Second, the small radius of
a black hole tends to imply a small magnetic flux.  On the other
hand, black holes may have a much shorter field correlation
time scale and may eject a much larger number of independent current
rings.

Suppose we consider the ejection of flux from a central black hole
in our galaxy (Hoyle 1969; Daly \& Loeb 1990).
A $10^6 M_{\sun}$ black hole, accreting at the Eddington limit
will have a mass accretion rate $\sim 10^{24}$ gm/s and a radius
of $3\times10^6$ km.  Consequently its characteristic flux
will be $\sim 10^{29}\ gauss\ cm^2$ or about 5000 times the flux
associated with any single protostar.  If the black hole magnetosphere
changes its magnetic structure in just a few light crossing times than
over the course of a few tens of millions of years it may eject as
many as $10^{13}$ current rings, as compared to the $\sim 10^{6.5}$ ejected
by a protostar.  We can conclude from this that the flux ejection
from a central black hole in our galaxy might be more effective by
a factor of order unity.  On the other hand, in addition
to the difficulty of mixing the ejected flux from a black hole into
the ISM it may also be that true the magnetic field of the black hole
magnetosphere is maintained by flux accreted from the surrounding
disk.  In this case the magnetic flux estimate may be too large by
a factor of $\alpha r/H$, which is probably of order unity, and a factor
which is a ratio of the correlation time scales for the poloidal field
in the accretion disk to light travel time across the black hole
magnetosphere.  The latter factor is difficult to estimate, but is
presumably a number substantially less than one.  Finally, it seems
doubtful that the formation of such a large black hole preceded
the epoch of star formation.

We might instead consider a population of solar mass black holes.
In this case the jets, although still narrowly focussed, are much
more likely to have mixed with the ISM.  The total number of such
objects is hard to estimate now, let alone
at earlier epochs, but if we express the total mass in such objects
in units of $10^6 M_{\sun}$ then since the radius of a black hole,
and its Eddington accretion rate, are proportional to its mass we find that the
total
seed flux  contributed
by such a population would be reduced from our previous estimate
by a factor of $ 10^{-3} M_6^{1/2}$, where $M_6$ is the total mass of
the population in units of $10^6 M_{\sun}$.  It would appear that
such a population is unable to produce a dominant contribution to
the galactic seed field.

I am happy to acknowledge helpful comments from Neal Evans and Robert Duncan.
This work was supported in part by NASA grant NAGW-2418.

\heading{REFERENCES}
\journal{Balbus, S., and Hawley, J.}{1991}{ApJ}{376}{214}
\journal{Basri, G., \& Marcy, G.W.}{1988}{ApJ}{330}{274}
\journal{Biermann, L.}{1950}{Zs. Naturforsch.}{5a}{65}
\journal{Bouvier, J., \& Bertout, C.}{1989}{Astron. \& Astrophys.}{211}{99}
\book{Chandrasekhar, S.}{1961}{Hydrodynamic and Hydromagnetic
Stability}{London}{Oxford University Press}
\journal{Daly, R.A., \& Loeb, A.}{1990}{ApJ}{364}{451}
\privcom{Dolgov, A.D., \& Rhie, S.H.}{1992}{preprint}
\infuture{Dolgov, A.D., \& Silk, J.}{1992}{Phys. Rev D}{in press}
\journal{Dopita, M.A., Schwartz, R.D., \& Evans, I.}{1982}{ApJ}{263}{L73}
\inbook{Fukui, Y., Iwata, T., Mizuno, A., Bally, J., \& Lane,
A.P.}{1992}{Protostars and Planets III}{E. H. Levy \& J. Lunine}{Tucson}{Univ.
of Arizona Press}{in press}
\journal{Hawley, J., and Balbus, S.}{1991}{ApJ}{376}{223}
\journal{Harrison, E.R.}{19070}{MNRAS}{147}{279}
\journal{Hayashi, C.}{1981}{Prog. Theor. Phys. Suppl.}{70}{35}
\journal{Hogan, C.J.}{1983}{Phys. Rev. Lett.}{51}{1488}
\journal{Hoyle, F.}{1969}{Nature}{223}{936}
\journal{Kronberg, P.P., \& Perry, J.J.}{1982}{ApJ}{263}{518}
\inbook{Kulsrud, R.M.}{1990}{Galactic and Intergalactic magnetic Fields}{R.
Beck, P.P. Kronberg, \& R. Wielebinski}{Dordrecht}{Kluwer}{219}
\infuture{Lazarian, A.}{1992}{AA}{in press}
\inbook{Levy, E.H. and Sonett, C.P.}{1978}{Protostars and Planets} {T.
Gehrels}{Tucson}{University of Arizona Press}{516}
\privcom{Liang, W.Z., Diamond, P.H., \& Vishniac, E.T.}{1992}{submitted to ApJ}
\journal{Mishustin, I.N. \& Ruzmaikin, A.A.}{1971}{JETP}{61}{441}
\journal{Mundt, R., \& Fried, J.W.}{1983}{ApJ}{274}{L83}
\journal{Parker, E.N.}{1955}{ApJ}{122}{293}
\journal{\samename}{1971}{ApJ}{163}{252}
\book {\samename}{1979}{Cosmical Magnetic Fields (their origin and
activity)}{Oxford}{Clarendon Press}
\journal{Piddington, J.H.}{1972}{Cosmic Electrodyn.}{3}{129}
\journal{Pudritz, R.E. \& Silk, J.}{1989}{ApJ}{342}{650}
\journal{Quashnock, J.M.,  Loeb, A., \& Spergel, D.N.}{1988}{ApJ}{254}{77}
\journal{Ratra, B.}{1992}{ApJ}{391}{L1}
\journal{Roxburgh, I.W.}{1966}{MNRAS}{132}{201}
\journal{Saar, S.H., \& Linsky, J.L.}{1985}{ApJ}{299}{L47}
\journal{Shakura, N.I., \& Sunyaev, R.A.}{1973}{Astr. Ap.}{24}{337}
\inbook{Stahler, S.W.}{1993}{Astrophysical Jets}{M. Livio, C.O'Dea, \& D.
Burgarella}{Cambridge}{Cambridge U. Press}{in press}
\book{Stix, M.}{1989}{The Sun: An Introduction}{Berlin: New
York}{Springer-Verlag}
\journal{Turner, M.S., \& Widrow, L.M.}{1988}{Phys. Rev}{D37}{2743}
\journal{Umebayashi, T.}{1983}{Prog. Theor. Phys.}{69}{480}
\journal{Umebayashi, T., and Nakano, T.}{1988}{Prog. Theor. Phys.
Suppl.}{96}{151}
\journal{Vachaspati, T.}{1991}{Phys. Lett.}{B265}{258}
\journal{Vainshtein, S.I., \& Cattaneo, F.}{1992}{ApJ}{393}{165}
\infuture{Vainshtein, S.I., Parker, E.N., \& Rosner, R.}{1993}{ApJ}{in press}
\journal{Vainshtein, S.I., \&Ruzmaikin, A.A.}{1971}{Astron. J. (SSSR)}{48}{902}
\journal{\samename}{1972}{Astron. J. (SSSR)}{49}{449}
\journal{Velikhov, E.P.}{1959}{Soviet JETP}{35}{1398}
\journal{Vilenkin, A., \& Leahy, D.A.}{1982}{ApJ}{254}{77}
\infuture{Vishniac, E.T., \& Diamond, P.H.}{1992}{ApJ}{in press}
\journal{Welter, G.L., Perry, J.J. \& Kronberg, P.P.}{1984}{ApJ}{279}{19}
\journal{Wolfe, A., Lanzetta, K.M., \& Oren A.L.}{1992}{ApJ}{338}{17}
\journal{Wood, J., and Mineshige, S.}{1989}{M.N.R.A.S.}{241}{259}
\vfill
\end